\documentclass[a4paper]{jpconf}
\usepackage{graphicx}
\usepackage{cite}
\begin{document}
\title{Differential top pair cross section and top anti-top plus jets Physics 
\footnote{Preprint number: WUB/13-02}}

\author{Malgorzata Worek}

\address{Theoretische Physik, Fachbereich C, Bergische Universit\"at 
Wuppertal, D-42097 Wuppertal, Germany}
\address{Institut f\"ur Theoretische Teilchenphysik und Kosmologie,
RWTH Aachen University, D-52056 Aachen, Germany}
\ead{worek@physik.rwth-aachen.de}

\begin{abstract}
A brief summary of the current status of the next-to-leading order QCD
corrections to top quark pair production  and the associated
production of $t\bar{t}$ with jet(s) in different  configurations,
{\it i.e.} with  one jet, two jets and another $t\bar{t}$ pair,  is
presented.

\end{abstract}

\section{Introduction}

By the end of 2012 a successful run of the Large Hadron  Collider
(LHC)   with  proton-proton collisions at  8 TeV  delivered $23.3$
fb$^{-1}$ of collected data.  This large available statistics has
opened a window on entirely new measurements of  more complex final
states. Reducing theoretical uncertainties for correct interpretation
of these data is among the highest priorities of the theoretical high
energy community. A theoretical accuracy at least at  next-to-leading
order (NLO) level is desirable and for most analyses even  demanded. 

The major purpose  of the LHC is  a deeper understanding of the
interactions  among fundamental constituents of matter. To this end,
the top quark   plays a special role. At the LHC it is copiously produced
via  strong interactions in top anti-top pairs, so its  production
cross section, decays and properties are studied with  high precision.
Apart from being studied as a signal process, the top quark constitutes
the main background  in analyzes of the  recently discovered
Standard Model (SM) like Higgs boson. Distinguished by its large mass,
the top quark is also potentially sensitive to physics beyond the SM
(BSM), which concerns among others searches for non-SM Higgs bosons and
supersymmetric particles.  In order to understand and control top
quark background processes to the SM and the BSM physics, precise
predictions for these reactions at the differential level are
indispensable.

In this contribution, a brief report on the recent activities in the
calculations of the NLO QCD  corrections to the differential top pair
cross section and  the associated production of $t\bar{t}$   with
jet(s) in different  configurations, {\it i.e.} with  one jet, two
jets and a $t\bar{t}$ pair, is given.

\section{Top Anti-Top }

The NLO QCD corrections to the production of the top anti-top pairs
followed by top decays in the so called narrow-width approximation
(NWA)  have already been  available for some time. In this approach
the  full amplitude is factorized into production of the unstable
particles and their subsequent decays neglecting non-resonant and
non-factorizable contributions to the amplitude.  The NWA
significantly simplifies calculations  of higher  order corrections on
the one hand at the same time  preserving gauge invariance.  First
calculations of the NLO QCD corrections to the  differential cross
section of top quark pair production in hadronic collisions  taking  only double
resonance contributions into account  and top and anti-top spin
degrees of freedom in the production and decay  have been presented in
\cite{Bernreuther:2001rq,Bernreuther:2004jv}.  Mixed QCD and 
electroweak NLO corrections have been also calculated see
e.g. \cite{Bernreuther:2010ny} and references therein.  In addition,
several charge asymmetries have been computed for the $t \bar{t}$
production  at the LHC and at the Tevatron \cite{Bernreuther:2012sx}.
Recently independent implementations of the  NLO QCD corrections to
top quark production and decay  at fully differential level with all
spin correlations appeared \cite{Melnikov:2009dn,Campbell:2012uf}.
Moreover, NLO QCD corrections matched to parton shower for heavy
flavor hadroproduction are obtainable  via publicly available
numerical programs like e.g. \textsc{MC@NLO} \cite{Frixione:2003ei} and
\textsc{Powheg} \cite{Frixione:2007nw}, however these programs do not
fully include spin  correlations through NLO QCD.

If resonant top production dominates, the NWA approach  is very  well
motivated. Nevertheless the accuracy of the NWA needs to be tested,
which  requires a full NLO QCD calculation of off-shell effects, i.e.
double-, single- and non-resonant top quark contributions of the order
${\cal O}(\alpha_s^3\alpha^4)$ need to be taken into account.  This
has been done by two independent groups that  have calculated NLO QCD
corrections to the production of top anti-top pairs including
interferences, off-shell effects, non-resonant contributions and  spin
correlations \cite{Denner:2010jp,Bevilacqua:2010qb,Denner:2012yc}.
This required the introduction of a complex-mass scheme for unstable
top quarks. Moreover, the intermediate W bosons were treated
off-shell.  These NLO calculations provide the most complete description of
top anti-top pair production with leptonic decays  of both  $W^{\pm}$
gauge bosons. For the inclusive cut selection finite-width effects  on
$\sigma_{t\bar{t}}$ have been found to be around  $1\%$ both at the
Tevatron and the LHC. In addition, the  asymmetries for the top quark
and the charged lepton  in top anti-top production at the Tevatron and
the LHC have been studied \cite{Bevilacqua:2010qb,Denner:2012yc}.

\section{Top Anti-Top Plus Jet}

For the $p\bar{p}$ and $pp$ collisions at the TeVatron and the LHC
a substantial number of events in the inclusive top anti-top sample is
accompanied  by an additional jet. Since top quarks are produced with
large energies and at high $p_T$, the probability for a top quark to
radiate gluons considerably increases. Depending on the $p_T$  of the
additional jet the fraction of events with an additional jet can
easily  be of the order of $30\%$ or more \cite{Dittmaier:2008uj}.
Sample  values for two  different  $p_T(j)$ cuts  are given in Table
\ref{tab1}, both for the TeVatron  and the LHC, where results for
$\sigma^{\rm NLO}_{t\bar{t}j}$ have been taken from \cite{Dittmaier:2008uj}, 
while  $\sigma^{\rm NLO}_{t\bar{t}}$ have been calculated with the help of  
\textsc{Top++} \cite{Czakon:2011xx}.
%
\begin{table}[h!]
\caption{\label{tab1}\it Fraction of events in the inclusive top
  anti-top sample that  is accompanied by an additional jet depending
  on the $p_T$  of the  additional jet at the TeVaron and the LHC.}
\vspace{0.4cm}
\begin{center}
  \begin{tabular}{lclc}
\br
 \textsc{TeVatron} &    
$\sigma^{\rm NLO}_{t\bar{t}j}/\sigma^{\rm NLO}_{t\bar{t}}$ 
&\textsc{LHC} & 
$\sigma^{\rm NLO}_{t\bar{t}j}/\sigma^{\rm NLO}_{t\bar{t}}$ \\
$\sqrt{s}=1.96$ TeV && $\sqrt{s}=14$ TeV&\\
\mr
$p_T \ge 20$ GeV  & $30\%$ & $p_T \ge 50$ GeV  & $47\%$\\
$p_T \ge 40$ GeV & $11\%$ & $p_T \ge 100$ GeV & $22\%$\\
\br
  \end{tabular}
\end{center}
\end{table}
%
The production of a top anti-top pair together with an additional jet
is therefore crucial for a more precise understanding of the topology
of top anti-top events. The $t\bar{t}j$ production is, however, also a
dominant background to various new physics searches. Among others,
the analysis of  the SM Higgs boson in the vector boson fusion process is
a prominent example. In this case, precise theoretical
predictions are indispensable.

NLO QCD corrections to the on-shell production of $t\bar{t}j$ have
been first  calculated in \cite{Dittmaier:2007wz} and subsequently
confirmed in \cite{Bevilacqua:2010ve,Melnikov:2010iu}. Furthermore in
\cite{Melnikov:2010iu} top quark decays at leading order (LO) in the NWA were
included.  Quite recently NLO QCD corrections to the production and
top quark decays in the NWA together with jet radiation in top
quark decays have been presented in \cite{Melnikov:2011qx}. Moreover,
spin correlations were preserved throughout the entire decay
chain. This state-of-the-art calculation of the NLO QCD corrections to
the $t\bar{t}j$ process has shown that  NLO QCD corrections and jet
radiation in  decays can lead to significant changes in shapes of
distributions. Therefore, they  need to be included for a fully
consistent description of top anti-top plus jet production.

First results for the top anti-top plus jet process at NLO combined
with  parton shower via \textsc{Powheg} method
\cite{Nason:2004rx,Alioli:2010xd} are also available
\cite{Kardos:2011qa,Alioli:2011as}.  However, in both cases only 
NLO QCD corrections to the on-shell production are calculated and 
LO decays in NWA are included   through the parton shower programs. 
Consequently  full spin correlations at NLO are omitted.

\section{Top Anti-Top Plus  Two Jets}

Even thought a fraction of events in the inclusive top anti-top sample
that  is accompanied by two additional jets is only at the few percent
level  \cite{Bevilacqua:2011aa}, see  Table \ref{tab2}, $t\bar{t}$
plus two jets is an important background for  the SM Higgs boson
studies at the LHC. Two noticeable examples include:
\begin{enumerate} 
\item  
$H \to WW^{*}\to \ell \ell' \nu \bar{\nu}$  where the Higgs boson is
  produced  via weak boson fusion \cite{Kauer:2000hi,Asai:2004ws},
\item  
$H\to b\bar{b}$ where the Higgs boson  is  produced via associated
  production with a $t\bar{t}$ pair \cite{Ball:2007zza,Aad:2009wy}.
\end{enumerate} 
In both cases  the invariant mass of the Higgs decay products cannot
be directly reconstructed.   Either  because of the two missing
neutrinos in the decay of the two $W$ gauge bosons or because the
$b\bar{b}$ pair can be chosen incorrectly within the complex
$W^{+}W^{-}b\bar{b}b\bar{b}$ final state.  In the latter case also the
b-tagging efficiency plays a crucial role since two b-jets can arise
from mistagged light jets.  Consequently, a very precise knowledge of
QCD backgrounds,  i.e. $t\bar{t}jj$ as well as $t\bar{t}b\bar{b}$ is
essential. Lately however,  a new strategy for the $Ht\bar{t}$ channel,  based
on a boosted Higgs boson and a boosted top quark,  has been explored
that should help to reduce complicated QCD backgrounds and resolve a
multi b-jet combinatorial problem \cite{Plehn:2009rk}. 
%
\begin{table}[h!]
\caption{\label{tab2}\it Fraction of events in the inclusive top
  anti-top sample that  is accompanied by two additional jets
  depending on the $p_T$  of the  additional jets at the TeVaron and
  the LHC.}
\vspace{0.4cm}
\begin{center}
  \begin{tabular}{lclc}
\br
 \textsc{TeVatron} &    
$\sigma^{\rm NLO}_{t\bar{t}jj}/\sigma^{\rm NLO}_{t\bar{t}}$ 
&\textsc{LHC} & 
$\sigma^{\rm NLO}_{t\bar{t}jj}/\sigma^{\rm NLO}_{t\bar{t}}$ \\
$\sqrt{s}=1.96$ TeV && $\sqrt{s}=7$ TeV&\\
\mr
$p_T \ge 20$ GeV  & $4\%$ & $p_T \ge 50$ GeV  & $6\%$\\
$p_T \ge 40$ GeV & $1\%$ & $p_T \ge 100$ GeV & $1\%$\\
\br
  \end{tabular}
\end{center}
\end{table}
%
Fortunately, the  NLO QCD corrections have already been calculated
for both   $t\bar{t}b\bar{b}$
\cite{Bredenstein:2009aj,Bevilacqua:2009zn,Bredenstein:2010rs,Worek:2011rd}
and $t\bar{t}jj$ \cite{Bevilacqua:2010ve,Bevilacqua:2011aa} background
processes at the TeVatron and the LHC. In the former case two
independent calculations exist and the per-mille level agreement
between them  have been obtained.  Nevertheless, due to complexity of
the NLO calculations for the $2\to 4$ processes   only  corrections to
the on-shell top production have been evaluated in all cases.
Furthermore, they have not yet been matched to the parton shower
programs. 

\section{Top Anti-Top Plus  Top Anti-Top}

\begin{figure}
\caption{\it  \label{fig1}  Differential cross section
  distributions as a function of the invariant mass of the $t\bar{t}t\bar{t}$
  system (left panel) and the  $t\bar{t}$ pair (right panel) for  $ pp \to t
  \bar{t}  t \bar{t} + X$ production at the LHC with $\sqrt{s}= 14
  ~\textnormal{TeV}$.  The dash-dotted (blue) curve corresponds to the LO,
  whereas the solid (red) one to the NLO result. The scale choice is $\mu_F =
  \mu_R = H_T/4$. The uncertainty bands depict scale variation. The lower
  panels display the differential $\cal K$ factor.}
\vspace{0.4cm}
\begin{center}
\includegraphics[width=0.45\textwidth]{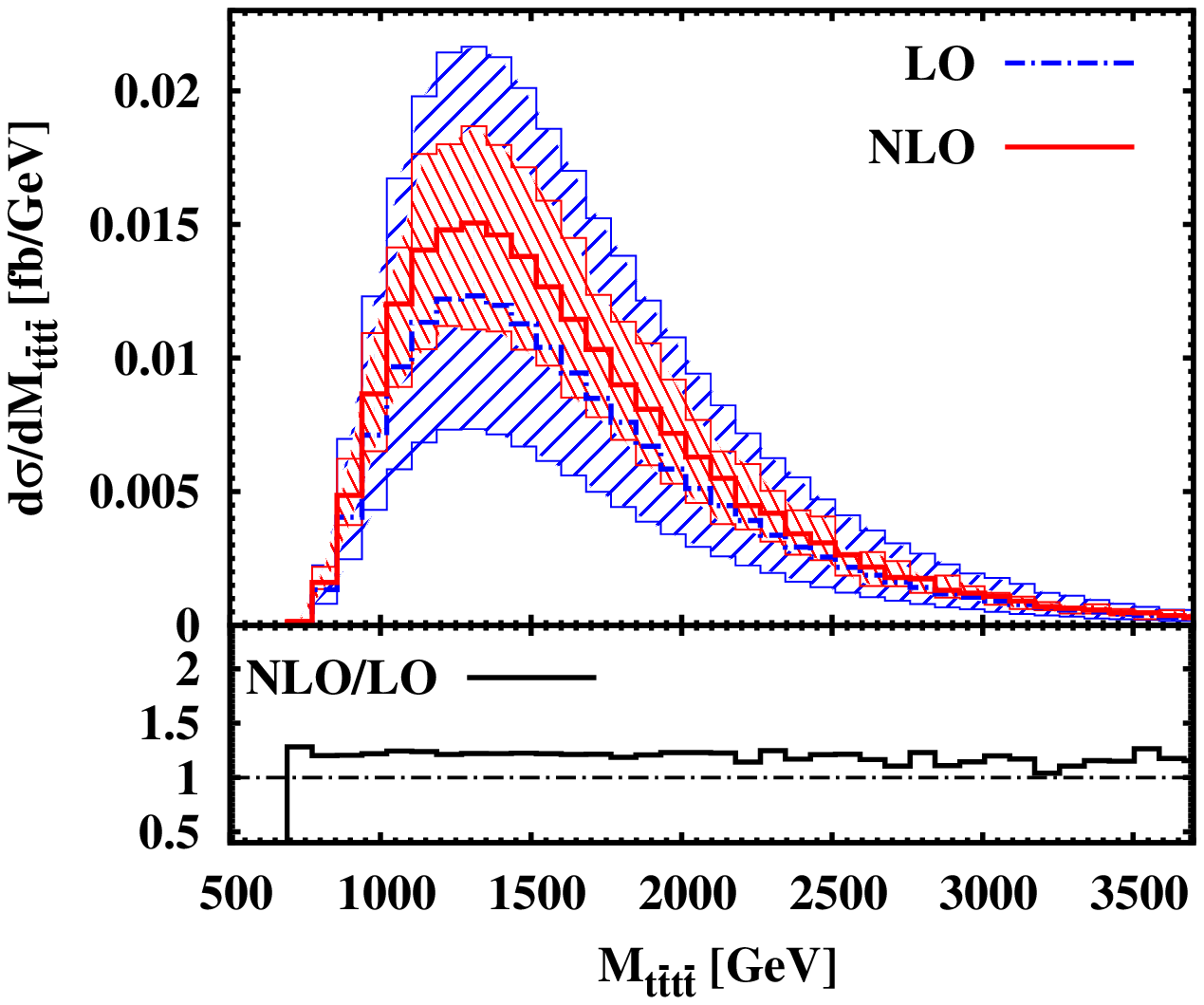}
\includegraphics[width=0.45\textwidth]{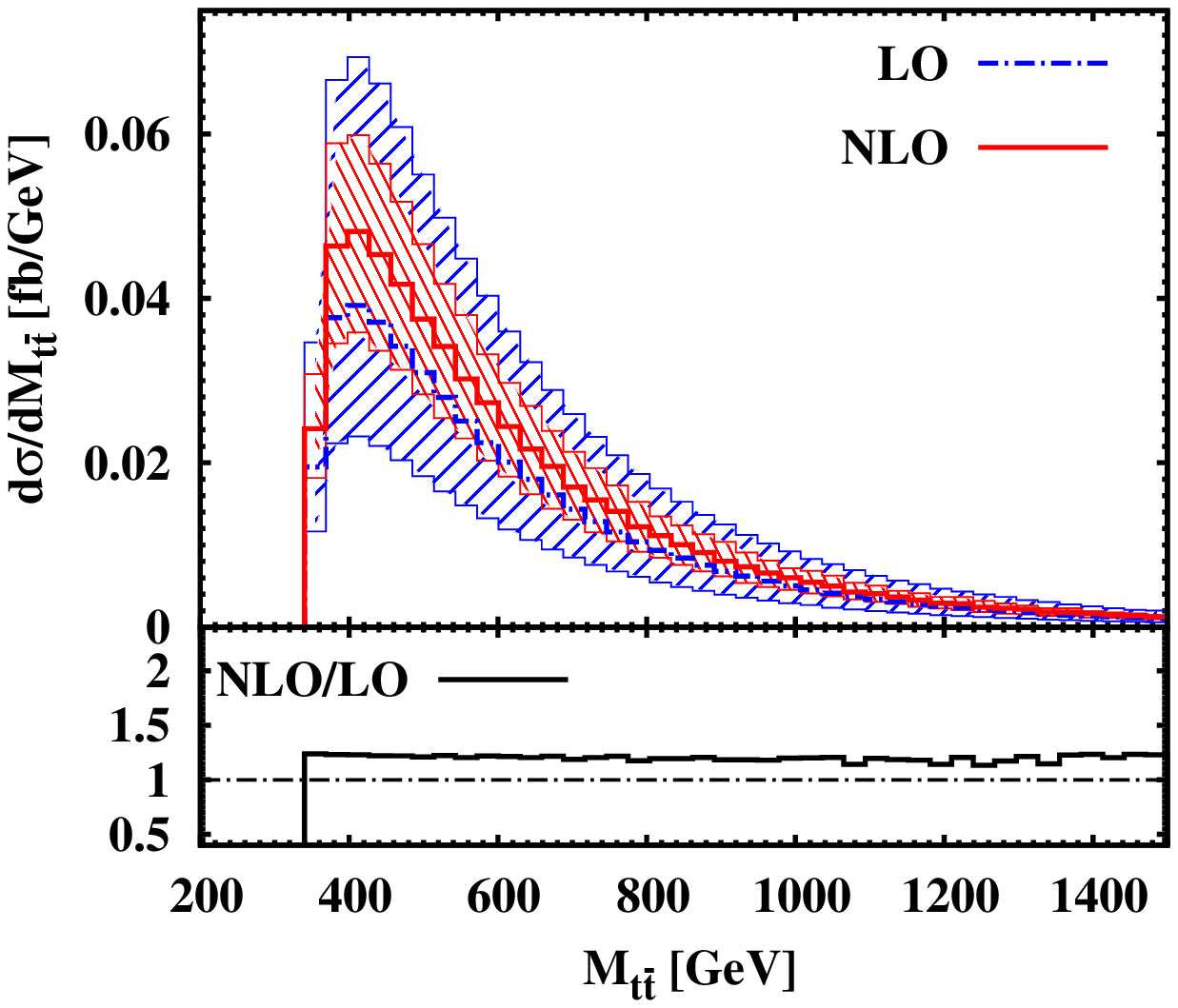}
\end{center}
\end{figure}
%
At the LHC the energy is sufficient to produce a four top final state at
a sensible rate, see  Table \ref{tab3}, where results obtained with
the \textsc{Helac-Dipoles} Monte Carlo program \cite{Czakon:2009ss},
for four different center of mass energies are given.
\begin{table}[h!]
\caption{\label{tab3}\it Inclusive leading order cross section in fb
  for  the $pp\to t\bar{t}t\bar{t}$ production at the LHC with
  $\sqrt{s}=7$ TeV, $\sqrt{s}=8$ TeV, $\sqrt{s}=13$ TeV and
  $\sqrt{s}=14$ TeV.  Results for the MSTW2008 LO PDF set are
  presented with the scale choice $\mu=2m_t$.  For the top quark mass
  the $m_t = 173.2$ GeV value is used.  Also given is a number of
  expected events assuming an integrated luminosity  of $5.6^{-1}$ fb,
  $23.3^{-1}$ fb, $100^{-1}$ fb and  $300^{-1}$ fb respectively.}
\vspace{0.4cm}
\begin{center}
  \begin{tabular}{lcc}
\br
\textsc{LHC} & 
$\sigma^{\rm LO}_{t\bar{t} t\bar{t}} $ [fb] & \textsc{Number of Events}\\
\mr
$\sqrt{s}=7$ TeV  & 0.624(1)	 & 3 \\
$\sqrt{s}=8$ TeV  &  1.173(3) & 27 \\
$\sqrt{s}=13$ TeV  & 9.08(3) & 908 \\
$\sqrt{s}=14$ TeV &  12.07(4) & 3621\\
\br
  \end{tabular}
\end{center}
\end{table}
%
The four top final state is an interesting channel to probe several
realizations  of BSM Physics. The most  noticeable models being 
\cite{Brooijmans:2010tn}
\begin{enumerate}
\item Higgs and top compositeness, 
\item new resonances from the Randall-Sudrum warped extra dimensions,
\item effective four-top quark interactions.
\end{enumerate}
In addition, $t\bar{t}t\bar{t}$ is a major background for many
processes  arising from supersymmetric extensions of the SM, among
others,  the production of a heavy Higgs boson or  long cascade decays
of colored new particles  like squarks or gluinos, see
e.g. \cite{Brooijmans:2010tn} and references therein.  Therefore, a precise
theoretical description of the four-top production rate in  the SM
might help to constrain new physics scenarios.

The NLO QCD corrections to the four top quark final state have been
recently  computed for on-shell tops \cite{Bevilacqua:2012em}.
Despite its relatively small cross section at NLO  of the order of 
\begin{equation}
\sigma^{NLO}_{t\bar{t}t\bar{t}}({\rm LHC}_{14{\rm TeV}}, m_t =173.2
~{\rm GeV, MSTW2008NLO}) = 17\pm 4  ~{\rm [scales]}  \pm 1 ~{\rm
  [PDF]} ~{\rm fb} 
\end{equation}
a theoretical control over $pp\to t\bar{t}t\bar{t}$ is relevant if
we take into account that typical predictions of various new physics
scenarios  are set in the range of 1-100 fb for $m_{new}$= 1-3 TeV
\cite{Brooijmans:2010tn}, where $m_{new}$ is a mass of the new heavy
particle or in general the energy scale that is associated with new
physics.  

In addition, a judicious choice of a dynamical scale has been
presented, that  allowed to obtain nearly constant ${\cal K}$-factors
in most differential  cross sections. Two examples, the invariant mass
of the $t\bar{t}t\bar{t}$ system  and the averaged  invariant mass of
the  $t\bar{t}$ pair are shown in Figure \ref{fig1}.

\section{Summary}
Driven by the LHC needs, a tremendous development in the NLO QCD
calculations for  top quark physics has recently been achieved.
Currently,  $2 \to 4$ processes are scrutinized at NLO. In many
cases dynamical scales that depend on the event structure have been
applied that  allow for a  better understanding of the high $p_T$
tails of distributions. For the top anti-top pair production with
leptonic decays the situation is the most satisfying, since 
complete off-shell and finite width effects of top quark have been
calculated at NLO.  Studies to match these calculations to the parton
shower programs via \textsc{Powheg} method  are currently ongoing
\cite{WWbb}.  For the semi-leptonic decay channel and more complex
final states  progress is still needed, an ultimate goal being a description of a 
fully realistic final state such as $W^{+}W^{-}b\bar{b}+X$ with $X =
j, jj, H, Z, W^{\pm}, \gamma$ matched to parton shower with higher
than LL accuracy.

Meanwhile a huge progress in  attempts to calculate the $t\bar{t}$
cross section beyond NLO has been attained.  An outstanding example
being a first genuine calculation of next-to-next-to-leading order
QCD corrections to top anti-top pair
\cite{Baernreuther:2012ws,Czakon:2012zr,Czakon:2012pz}.

\ack We would like to thank the organizers of the  5th International
Workshop on Top Quark Physics, TOP2012, for a kind invitation and a
very pleasant atmosphere during the conference.

\end{document}